\documentclass{article}

\usepackage{epsfig}
\usepackage{graphics}
\usepackage{graphicx}
\usepackage[centertags]{amsmath}
\usepackage{amsfonts}
\usepackage{amsthm}
\usepackage{amssymb}

\usepackage{url}
\usepackage{subfigure}
\usepackage{hyperref}

\begin{document}

\title{Brief Notes and History Computing in Mexico during 50 years\footnote{Genaro J. Mart{\'i}nez, Juan C. Seck-Tuoh-Mora, Sergio V. Chapa-Vergara and Christian Lemaitre (2019) Brief notes and history computing in Mexico during 50 years, {\em International Journal of Parallel, Emergent and Distributed Systems}. DOI \url{https://doi.org/10.1080/17445760.2019.1608990}}}

\author{Genaro J. Mart{\'i}nez$^{1,2}$, Juan C. Seck-Tuoh-Mora$^{3}$ \\ Sergio V. Chapa-Vergara$^{4}$, Christian Lemaitre $^{5}$}


\maketitle

\begin{centering}
$^1$ Escuela Superior de C\'omputo, Instituto Polit\'ecnico Nacional, M\'exico. \\
$^2$ Unconventional Computing Lab, University of the West of England, \\ Bristol, United Kingdom. \\
$^3$ \'Area Acad\'emica de Ingenier{\'i}a, Universidad Aut\'onoma del Estado de Hidalgo, Pachuca, M\'exico. \\
$^4$ Centro de Investigaciones y Estudios Avanzados del Instituto Poli\'ecnico Nacional, M\'exico, D.F. \\ 
$^5$ Universidad Aut\'onoma Metropolitona, Unidad Cuajimalpa, M\'exico, D.F. \\ 
\end{centering}

\begin{figure}[th]
\centerline{\includegraphics[width=3.5in]{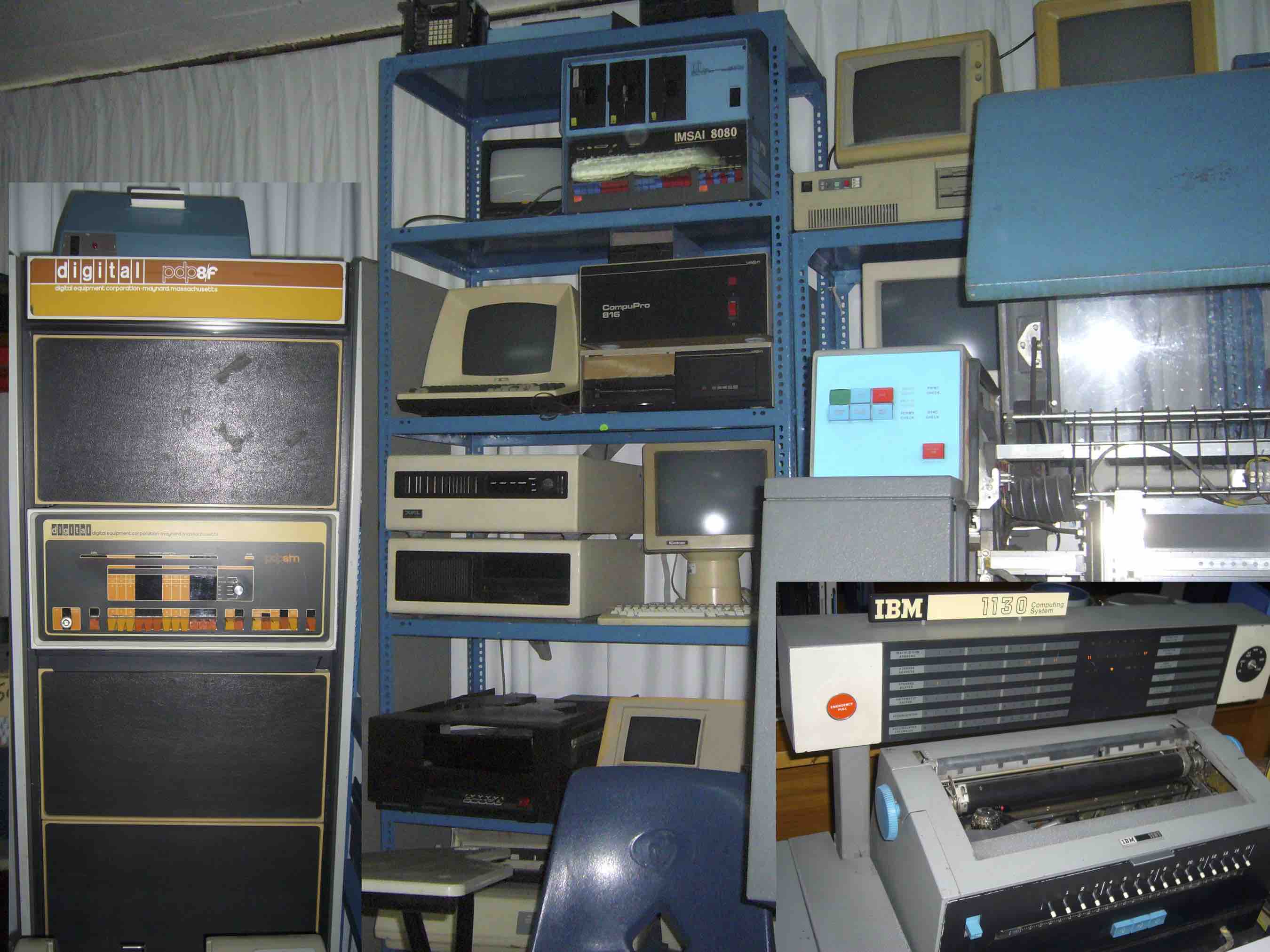}}
\caption{McIntosh's computer museum, Puebla, Mexico.}
\end{figure}

The history of computing in Mexico can not be thought without the name of Prof. Harold V. McIntosh (1929-2015) \cite{kn:Cis15, kn:McJ15, kn:Mart16}. For almost 50 years, in Mexico he contributed to the development of computer science with wide international recognition. Approximately in 1964, McIntosh began working in the Physics Department of the Advanced Studies Center (CIEA) of the National Polytechnic Institute (IPN), now called CINVESTAV. In 1965, at the National Center of Calculus (CeNaC), he was a founding member of the Master in Computing, first in Latin America. With the support of Mario Baez Camargo and Enrique Melrose, McIntosh continues his research of Martin-Baltimore Computer Center and University of Florida at IBM 709. 

An excellent quality of McIntosh was his vocation as a professor, which meant a great recognition by his students. For this reason, he was strongly supported by Roberto Mendiola, head of the School of Physics and Mathematics (ESFM), where McIntosh was a professor from 1966 to 1975 and coordinator of the Academy of Applied Mathematics. In the National Institute of Nuclear Energy (INEN) from 1971 to 1975 he research in the Salazar Computer Center. Under the administration of the physicist Juan Jos\'e Ort{\'i}z Amezcua, the collaboration of Tom's Brody and a group of young people, programs and software were developed in a network of computers such as PDP-10, PDP-11 and PDP-15 connected between them. In this way, computer research at the INEN represented the value of teaching in the ESFM, as one of McIntosh's greatest legacies in Mexico. 

A change of employment of the computer group of the INEN to the Autonomous University of Puebla (UAP) coincided with the commercialization of microcomputers. In 1976, Luis Rivera Terrazas, head of the UAP, contracted to McIntosh and a group of 12 INEN researchers to consolidate the Bachelor's Degree in Computer Science and form a computer research project. Thus arose the Department of Application of Microcomputers in Puebla, with the intention of building microcomputers with microprocessors such as: Motorola 6800, WD1600, Intel 8080, IMSAID, and others. This was the opportunity to request the background of the compiler REC/MARKOV to develop operating systems for microcomputers \cite{kn:Mc86}. 

McIntosh's programming experience earned at the Martin Company in the US UU. Writing FLT and MBLISP \cite{kn:Mc62} was the antecedent to start a computer project in Mexico in the IBM 709. LISP was the invention of John McCarthy on a theoretical concept like the Lambda calculus of Church and the concept of Turing of a universal machine In this way LISP was the representative of symbolic computing. In McIntosh's research in Quantum Chemistry, it was the main motive for the development of MBLISP for symbolic differentiation, the Poisson calculation or the commutator brackets and, in general, computational group-theoretical methods. 

Mathematical logic was one of the main axes of the courses of the Academy of Applied Mathematics at the ESFM, offered jointly with CeNaC in the master's program in computer science. The courses were held to alternate between a practical version dedicated to programming in the PDP-8 and a theoretical vision of Boolean Algebra, the theory of automata and the elements of computability theory. In their two roles, they played the development of programming skills and learning the fundamentals of computer theory. First, starting with the physical description of the PDP-8, to analyze the editor and the assembler, but the best of all was the DDT (Dynamic Debugging Technique) program. The DDT programs in the small DEC computer were very important for the student to have complete control of the structure and operation of this program, reason for the REC/DDT projects for the PDP-8, PDP-15, PDP-12 and PDP-11. On the other hand, the teaching of infinite machines and the theory of computability, the Turing machine, the Post systems and the Markov algorithms \cite{kn:Mc67}. 

REC (Regular Expression Compiler) was a relevant programming language in the history of the Academy of Applied Mathematics of the ESFM [8]. With a pedagogical desire, it was used as a language model of a finite state machine to process four control symbols based on regular expressions. REC has the characteristic of dealing with recursively defined data structures, typical of many symbol manipulation tasks and, characteristic of REC compiler writing. The particular version of REC arises from the choice of collections of arithmetic operators with the REC / arithmetic version similar to APL. However, the flexibility to create other compilers results in REC for special purposes: REC/DDT for debugging programs, REC/MA for controlling the multichannel analyzer and REC/Visual for visual control [9]. Indeed, REC and CONVERT \cite{kn:GM66} were cited notoriously by Marvin Minsky in his historic book ``Computation: Finite and Infinite Machines'' \cite{kn:Mins67}. 

In the direction of processing the Markov algorithms, McIntosh proposed the development of REC/Markov for the PDP-10 \cite{kn:Gar75}. An imitation text editing language or as a supplement to TECO; editor, available on the PDP-10's timeshare system. Originally, seven operators were provided, enough to program Markov algorithms studied in symbolic logic courses. Markov algorithms are very fundamental for the theory of computation, but they are not practical for text editing because they only allow the substitution of fixed character strings after a search from the beginning of the text. By allowing additional operations and the handling capabilities of adjacent files, a highly useful and efficient text editor emerges. In the Nuclear Center of Salazar (INEN) it was used for several years as a partner of TECO and calling routines of Fortran. 

By tradition, the numerical analysis in the Academy of Applied Mathematics had as courses the diagonalization of matrices and numerical integration of ordinary differential equations, with emphasis on systems of linear differential equations. The important thing was the research in theoretical-computational methods for groups applied to the solution of the wave equation of quantum mechanics. When many of the students were connected to the Mexican Petroleum Institute, there was a lot of interest in numerical analysis with physical-chemical applications. From the point of view of programming, Mc developed in the IBM 1130 and PDP-10, gradually a family of matrix arithmetic packages: MATRIX, PENTA and TRI, together with the integration of Runge-Kutta SERO and SECO and, TWOC for Hamiltonian mechanics. From the theoretical point of view, they dealt with the matrizante, its factorization and, the behavior of its eigenvalues and its eigenvectors. For the application in quantum mechanics, spectral density and Weyl's theory of singular equations \cite{kn:ShVMc15}. 

In the INEN, one of the main objectives of the purchase of its PDP-10 was to carry out a radiological study of mineral deposits in Mexico, for which it was possible to convince the administration to start a new computer graphics project. The fulfillment of this obligation led mainly to initiate programs of data reduction and contour graphics. PLOT was a collection of subroutines for the CalCom tracer, which consists of most contour programs and hidden line subroutines. The design programs were divided into several more specific areas, reaching a wide range of applications. In this way, PLOT \cite{kn:Mc74} was very successful in the INEN, promoting a wide dissemination by students of the ESFM with its installation in a great majority of the computers in Mexico City, especially in Petroleos Mexicanos (PEMEX).  Aside from developing an abstract plotter program, McIntosh paid considerable attention to applications, such as the collection of programs to draw spheres GEOM. In this way, advances in the PLOT package allowed the Academy of Applied Mathematics to advance in the field of projective geometry, with a package such as PHOC (Plane Homogeneous Coordinates) and SHOC (Space Homogeneous Coordinates).\footnote{Some of these original papers can be found in McIntosh's webpage \url{http://delta.cs.cinvestav.mx/~mcintosh/cellularautomata/Welcome.html}} 

The first electronic computer in Mexico was an IBM 650 installed in the National Autonomous University of Mexico (UNAM), in June 1958. The team responsible of this project was a selected group of researchers in Engineering, Physics and Mathematics working in key roles at UNAM: Nabor Carrillo (Rector); Alberto Barajas (Sciences’ Provost); Carlos Graef (Dean of the Faculty of Science); and Sergio Beltran, leader of a research group at the Faculty of Engineering, who was named the first director of the UNAM Computer Center. In those days, very few people were aware of how to run an electronic computer, only some few researchers in Physics and astronomy had some experience using computer programs to do part of their PhD work in US Universities. 

Beltran showed a great creativity and enthusiasm organizing the Computing Center with students of engineering and physics, training them with the support of IBM, and organizing an annual colloquium for researchers and students with some of the top researchers at the time. From 1959 to 1962, Beltran organized four international colloquiums of Computer Science, among the invited key speakers we find, Alan Perlis, McCarthy, Minsky, McIntosh and Niklaus Wirth \cite{kn:Sor85}. 

During the 60’s years, other Universities like the IPN, and the private institution, the Technological Institute of Monterrey (ITESM), installed their own computer centers. The training programs of the academic computing centers added to those of the computer industry, allowed the first government institutions and some few private enterprises to install their own computer centers. 

By 1968, the first generation of Mexican students with a PhD in computer Sciences came back to Mexico: Renato Iturriaga, Adolfo Guzm\'an, Enrique Calder\'on, and Mario Magidin. They were the result of a successful policy in developing scientific infrastructure in many third world countries at the time, through an economic effort sending brilliant students abroad with a scholarship to get their PhD hopping they shall come back to reinforce the advance of scientific capacities of the country. This mechanism has been the only one used for almost four decades, until the first high standing doctorate programs began to produce the first PhD students in computer science in Mexico, around 2000 year. 

We must insist that Mexico, as many other countries, start to build his computing scientific and technological infrastructure from scratch. By the early 1980’s there were two main research groups with around 22 researchers each, one at the UNAM and the other at the UAP. Besides those groups there were several smaller ones scattered in different universities. It was a promising situation. Unfortunately, Mexico and other Latin American countries felt into a huge dept crisis in 1982 with very serious consequences for their finances, and particularly for their scientific institutions. By 1984, the two main computing research groups shrank from 22 to 4 researchers each as a result of the impact of the economic crisis on the higher education and scientific systems. 

This new economic model prevailing since then does not put the national scientific and technological development as a key economic variable. The needs of new technologies for government and industry are fulfilled by buying it abroad. Nevertheless, the field has being growing steadily since the mid-1980s due to two main forces: the support of the universities and the tenacity and excellent work of the researchers in computer science. The absence of strong computer science funding boosts the coordination and networking efforts of researchers. At the national level, different communities create their scientific associations. Just after the first strike of the economic crisis, several researchers in Artificial Intelligence decided to organize in 1986 the Mexican Association for Artificial Intelligence (SMIA), in order to organize and boot the research in the area. In 1988, with the national associations for Artificial Intelligence of Spain and Portugal, they create the Iberoamerican Association for Artificial Intelligence (IBERAMIA), and its biennial conference. In 1995 the Mexican Association of Computing Science was created. Since then other areas had launched their own associations: Robotics, Human Machine Interaction, Natural Language Processing, etc. This effort of grouping had led to the creation in 2015 of the Mexican Academy of Computing (AMEXCOMP) who is officially recognized and supported by the National Council of Science and Technology (CONACyT). This networking effort of the Mexican computing science community includes too the Mexican researcher's diaspora mainly in the U.S., Canada and Europe. 

During 90s and 2000 years, Mexico has a lot of contributions in different areas of computer science. In this stage, we can talk about a second stage of contributions inspired and motivated by McIntosh. This is in the theory of cellular automata. Along of all this time, McIntosh established relevant contact with prominent and historic contemporary professors, such as: Andrew Adamatzky, Matthew Cook, Edward Fredkin, Kenichi Morita, Stephen Wolfram, Jarkko Kari, Burton Voorhees and Andrew Wuensche, just to mention some.

The development of the called unconventional computing and/or natural computing scientific areas have increased significantly. Particularly, the international conferences specialized and dedicated to promote all these advances is the ``International Conference on Unconventional Computation and Natural Computation (UCNC)'' and the publication of these subject is published in specialized journals as: the International Journal of Unconventional Computing\footnote{\url{https://www.oldcitypublishing.com/journals/ijuc-home/}}, the Natural Computing journal\footnote{\url{https://link.springer.com/journal/11047}}, the IEEE Transactions on Evolutionary Computation\footnote{\url{https://ieeexplore.ieee.org/xpl/RecentIssue.jsp?punumber=4235}} and the ACM journal on Emerging Technologies in Computing.\footnote{\url{https://jetc.acm.org}}. Indeed several of these models developed in unconventional computing are designed with cellular automata \cite{kn:Hey98, kn:Ada17a, kn:Ada17b, kn:Zen12}.

From von Neumann era, the idea of supercomputing in non-linear media was a central problem in cellular automata theory \cite{kn:von66}. This concept and proposal are explored on diverse ways, such as: irreducible signals, gliders, self-mobile localizations, particles, waves, or molecules \cite{kn:Toff98, kn:Ada02}. The power of computation in cellular automata has been researched mainly on the name of `complex rules' \cite{kn:Wolf88, kn:MSZ13}. McIntosh had discusses several times in his classroom which von Neumann have captured the essence of massive computation and the cellular automata is a natural consequence to understand the power of computation both conventional and unconventional. Recurrently, it is a very long discussion about of von Neumann architecture versus non-von Neumann architectures into computer science community. McIntosh condense all his contribution in cellular automata with his book ``One-Dimensional Cellular Automata'' published at United Kingdom for the 2009 year \cite{kn:Mc09}. An excellent book applying several of McIntosh's analysis was wrote by Burton Voorhees in ``Computational Analysis of One-dimensional Cellular Automata'' \cite{kn:Voor96}. In this history, McIntosh had some personal meetings in his office with Fredkin and in other time with Morita and Wuensche both in 2011 year.

One of the most relevant subjects which McIntosh dedicated in his last years was to the problem of reversible cellular automata, a classical research field because these systems can conserve the initial information of the automata. This property has been of great interest for theoretical reasons to understand how information is conserved and can be recovered in the evolution of a discrete system, and for its computational implications.

McIntosh began his studies in reversible cellular automata in the early '90s, in these studies had the vision to include undergraduate and graduate students through the Research Summers organized by the Mexican Academy of Sciences (AMS). In these summers, the group of students in charge of McIntosh had the opportunity to assimilate the knowledge about the graphical tools used to analyze the one-dimensional cellular automata, such as: de Bruijn, pair, and subset diagrams. So, observe which properties of reversibility were characterized with these tools.\footnote{Research Summer guided by McIntosh at UAP. \url{http://uncomp.uwe.ac.uk/genaro/Papers/Veranos_McIntosh.html}}

These studies were enriched with the revision of the works of Gustav Hedlund \cite{kn:H69}, Masakazu Nasu \cite{kn:Nasu77} and Kari \cite{kn:Kari96} that from different perspectives allowed to explain and characterize the reversible behavior. It should be noted that this was a unique work in Mexico led by McIntosh in the Department of Application of Microcomputers of the UAP. The systematic and theoretical study of the reversible cellular automata led to the graduation of several undergraduate students and postgraduate studies from UNAM and CINVESTAV, who carried out their theses on these topics.\footnote{Special issue dedicated to McIntosh in the Journal of Cellular Automata at 2008. Discrete Tools in Cellular Automata \url{https://www.oldcitypublishing.com/journals/jca-home/jca-issue-contents/jca-volume-3-number-3-2008/}} This work in the investigation of the calculation of reversible automata allowed to have direct contact with international professors such as Wolfram, David Hillman who were also working on similar algorithms \cite{kn:Hillman91}.

Another important point that was studied with McIntosh was the question of giving a limit for the maximum length of the minimum neighborhood required to have a reversible behavior in evolution rule of one-dimensional cellular automata. Czeizler and Kari solved this problem \cite{kn:Czeizler04} \cite{kn:Czeizler05}, which allowed to have direct communication with them that was reflected in a visit to Finland as part of a {\it Workshop on Discrete Models for Complex Systems} held in 2004.

Derived from the studies on de Bruijn diagrams and subsets, and in talks with McIntosh about its possible applications, the conceptualization and application of an algorithm for the calculation of preimages in one-dimensional cellular automata \cite{kn:Mora04} was reached. The study of the properties of reversible cellular automata using graphical tools such as de Bruijn diagrams and Welch diagrams gave rise to new ways of calculating several systems using symbolic dynamics tools, such as the use of full-shift partitions reported in \cite{kn:Mora04b}

The work of McIntosh continues to have a relevant influence in later studies beyond the understanding of the properties of reversible automata and the use of various graphical tools. This influence has generated other types of results, such as the classification of behaviors in elementary cellular automata \cite{kn:Seck14}, the calculation of reversible automatons applying random block permutations \cite{kn:Seck05}, a direct use of the Welch indices for the random generation of reversible rules \cite{kn:Seck17} and applications of cellular automata to simulate manufacturing systems based on de Bruijn diagrams \cite{kn:Barragan18}.

In 2011 a cellular automata collider was presented in the paper titled ``Cellular automaton supercolliders'' \cite{kn:MAS11} where a virtual cellular automata collider can simply a computation. Derived as a large research started at Department of Application of Microcomputers of the UAP lead by McIntosh from 1998. In this time a relevant result in cellular automata theory was done. In 1998, Mathew Cook proofed which the elementary cellular automaton rule 110 is universal, the result was presented in a special workshop organized by the Santa Fe Institute in New Mexico, USA. Cook invented a novel finite machine, a cyclic tag system adapted to work on millions of millions of cells with dozens of particles coded, controlled and synchronized in the evolution of the one-dimensional cellular automata rule 110 \cite{kn:Cook04}. Two years later this result was published by Wolfram in his book ``A New Kind of Science'' \cite{kn:Wolf02}. By the way, McIntosh and students offering a different point of view to understand the dynamics of this automaton. McIntosh propose which the problem of rule 110 is a tiling problem \cite{kn:Mc99}. The number of results captured the attention of Wolfram and a meeting was celebrated in one of his events named NKS Conferences in Boston Massachusetts in USA by 2004 year.\footnote{A Way to Construct Complex Configurations in Rule 110. \url{https://www.wolframscience.com/conference/2004/presentations/HTMLLinks/index_43.html}}

This cellular automata collider was developed during two years in collaboration with the Nuclear Science Institute (ICN) of the UNAM, the Department of Application of Microcomputers of the UAP, and the Unconventional Computing Lab (UCL) at UWE in Bristol, United Kingdom. So, outstanding by the Massachussets Institute of Technology publication with the article titled ``Computer Scientists Build Cellular Automaton Supercollider''.\footnote{Computer Scientists Build Cellular Automaton Supercollider. \url{https://www.technologyreview.com/s/424096/computer-scientists-build-cellular-automaton-supercollider/}} In this stage, six years later finally a full simulation of this virtual collider was published the last year in a specialized book ``Advances in Unconventional Computing: Volume I Theory'' with the paper ``A Computation in a Cellular Automaton Collider Rule 110'' \cite{kn:MAM17, kn:MAM15}. At the same time, this reference was aggregated to the European Organization for Nuclear Research (CERN) digital library.\footnote{\url{http://cds.cern.ch/record/2216884}}

A cellular automaton collider is a finite state machine build of rings of one-dimensional cellular automata. The infinite computation which was designed originally by Cook is compressed in a circular machine coding package of particles by regular expressions.

In the late 70s Edward Fredkin and Tommaso Toffoli proposed a concept of computation based on ballistic interactions between quanta of information that are represented by abstract particles~\cite{kn:FT82}. The Boolean states of logical variables are represented by balls or atoms, which preserve their identity when they collide with each other. Fredkin, Toffoli and Norman Margolus developed a billiard-ball model of computation, with underpinning mechanics of elastically colliding balls and mirrors reflecting the balls' trajectories. Margolus proposed a special class of cellular automata which implements the billiard-ball model~\cite{kn:MTV86} named partitioned cellular automata exhibited computational universality because they simulated Fredkin gates via collision of soft spheres~\cite{kn:Ada02}. Also, the construction of this cellular automata collider consider previous results about circular machines designed by Michael Arbib, Manfred Kudlek, and Yurii Rogozhin.

Cellular automata collider use cyclotrons to explore large computational spaces where exact structures of particles are not relevant but only the interactions between the particles. There we can represent the particles as points and trains of particles as sequences of points. The cellular automata collider is a viable prototype of a collision-based computing device. It well compliments existing models of computing circuits  based on particle collisions. Thus, the simulated collider have just thousands of cells not millions.

Actually Mexico has very few research groups dedicated to unconventional or natural computing. However, an interesting group working about unconventional computing for several years following the tradition of summer school in Mexico educating dozens of Mexican students during the last decade. It is impulsed by Salvador El{\'i}as Venegas Andraca from ITESM, forming students and research groups particularly in quantum computing \cite{kn:Vene18}.

In 2017, the Faculty of Computation in UAP organized the ``Celebration of late Prof. Harold V. McIntosh achievements'' with the participation of several conferences on-line given by professors Leon Chua, Morita, Adamatzky and Wolfram.\footnote{\url{http://uncomp.uwe.ac.uk/HVM/}} By the way, in 2018 we celebrate the 60 anniversary of computation in Mexico.\footnote{60 a\~nos de la computaci\'on en M\'exico y la influencia de Harold V. McIntosh. \url{http://www.comunidad.escom.ipn.mx/ALIROB/Events/60CompMxHVM25ESCOM/}}

In 2018 year, it was done the construction of the first Turing machine in Mexico, the first robotic Turing machine internationally: the Cubelet-LEGO Turing machine (CULET) \cite{kn:FZM19}. This Turing machine define its construction with Cubelets robots and helped with LEGO pieces. The machine works on a fixed type and the head is controlled by the concatenation of several robots. This way, you can program any 2-symbol Turing machine in CULET. Actually, the machine can simulate the behaviour of a universal Turing machine simulating the universal elementary cellular automaton rule 110\footnote{\url{https://youtu.be/GIQDA5Gnxkc}} and a Turing machine which double the number of ones.\footnote{\url{https://youtu.be/CrUrKtIw34w}} The machine was constructed in the Artificial Life Robotics Lab (ALIROB)\footnote{ALIROB \url{http://www.comunidad.escom.ipn.mx/ALIROB/}} and the Computer Science Laboratory (LCCOMP)\footnote{LCCOMP \url{http://uncomp.uwe.ac.uk/LCCOMP/}} at the Escuela Superior de C\'omputo (ESCOM) of the IPN in collaboration with the Unconventional Computing Lab (UCL) of UWE. The robotic Turing machine was presented this year in the ``2019 International Conference on Artificial Life and Robotics'' (ICAROB2019), Oita in Japan. By the way, the founder of Cubelets robots, Eric Schweikardt, comments about this maquine in the forum of Modular Robotics with a post titled ``Can you make a computer out of Cubelets?''\footnote{\url{https://www.modrobotics.com/2019/01/28/can-you-make-a-computer-out-of-cubelets/}} The machine was an old project which was worked during more than two years in an international cooperation between IPN and UWE.


\end{document}